# Pressure effects in the triangular layered cobaltites $Na_xCoO_2$

G. Garbarino

Centre de Recherches sur les Très Basses Températures, C.N.R.S., BP 166 cedex 09,

38042 Grenoble

M. Monteverde[1]

Centre de Recherches sur les Très Basses Températures, C.N.R.S., BP 166 cedex 09,

38042 Grenoble and

Laboratorio de Bajas Temperaturas, Departamento de Física, Universidad de Buenos

Aires, Ciudad Universitaria, PAB I, 1428 Buenos Aires, Argentina

M. Núñez-Regueiro

Centre de Recherches sur les Très Basses Températures, C.N.R.S., BP 166 cedex 09,

38042 Grenoble, France

C. Acha

Laboratorio de Bajas Temperaturas, Departamento de Física, Universidad de Buenos

Aires, Ciudad Universitaria, PAB I, 1428 Buenos Aires, Argentina

M.L. Foo and R.J. Cava

Department of Chemistry and Materials Institut, Princeton University, NJ 08544,

USA

---

[1] Present address *Service de Physique de l'Etat Condensé (CNRS URA 2464), DSM/DRECAM/SPEC, CEASaclay, 91191 Gif sur Yvette Cedex, France*

We have measured transport properties as a function of temperature and pressure up to $30 GPa$ in the $Na_xCoO_2$ system. For the $x = 0.5$ sample the transition temperature at 53K increases with pressure, while paradoxically the sample passes from an insulating to a metallic ground state. A similar transition is observed in the $x = 0.31$ sample under pressure. Compression on the $x = 0.75$ sample transforms the sample from a metallic to an insulating state. We discuss our results in terms of interactions between band structure effects and $Na^+$ order.

72.80.Ga, 61.50.Ks, 62.50.+p

The layered cobalt oxides of the type $Na_xCoO_2$ have been studied in recent years due to their unexpected high thermoelectric power[1] and the possibility of frustration of a one half spin with anti-ferromagnetic interactions in a triangular lattice[2]. The interest increased when it was reported that by intercalating $Na_{0.3}CoO_2$ with water molecules the material could be rendered superconducting[3] at $T_c^{SC} \sim 5K$. On the other hand, the studies up to date have yielded a phase diagram in sodium concentration with properties essentially inverse to those envisioned[4]. Curie-Weiss behavior due to localized states is observed in concentrations near the band insulator point $(x = 1)$, while more metallic delocalized dependences are measured near the Mott-Hubbard insulating concentration $(x = 0)$. To the already complicated picture, the evidence of the presence of sodium ion ordering[5] at almost all concentrations[6] has added another crucial parameter to the problem. In spite of the large number of papers in the subject[7,8,9,10] a full understanding of the electronic structure, or of the importance of charge ordering in transport properties and in superconductivity, has yet to be achieved.

We have performed a detailed pressure and temperature study of transport properties for three different $Na$ concentrations, $x = 0.75$, $x = 0.31$ and $x = 0.5$. For the compound $x = 0.75$ we show that pressure causes an increase of the low temperature resistance. For the $x = 0.5$ compound we observe that the transition temperatures increase with pressure, while their effect on the transport properties disappears, down to a metallic state at the highest pressures. While for the $x = 0.31$ samples we observe that pressure triggers a transition similar to the one reported for the $x = 0.5$ compound.

The measured samples are powders from the same type of preparation as those of Ref. 4. $Na_{0.77}CoO_2$ was made by solid-state reaction of stoichiometric amounts of $Na_2CO_3$ and $Co_3O_4$ in oxygen at 800 C. Sodium de-intercalation was then carried out by treatment of samples in solutions obtained by dissolving $I_2$ (0.2 M, 0.04 M) or $Br_2$ (1.0 M) in acetonitrile. After magnetic stirring for five days at ambient temperature, they were washed with copious amounts of acetonirile and multiple samples were tested by the Inductively Coupled Plasma - Atomic Emission Spectrometer (ICP-AES) method to determine the $Na$ content.

The electrical resistance measurements were performed in a sintered diamond or carbon tungsten Bridgman anvil apparatus using a pyrophillite gasket and two steatite disks as the pressure medium[11]. The Cu-Be device that locked the anvils could be cycled between 4.2K and 300K in a sealed dewar. Three different pressure ranges: $0.7 - 8 GPa$, $1.4 - 22 GPa$ and $4 - 30 GPa$ were used on several powder samples of the $x = 0.31$, 0.5 and 0.75 compounds from the same batches. Due to the powder nature of our samples no ambient pressure resistivity measurement was possible.

a) $Na_{0.5}CoO_2$. On Fig. 1, we show the variation of the electrical resistance temperature dependence with pressure for the $x = 0.5$ concentration. At the lowest pressures, we observe the transition to an insulating state as previously reported in monocrystal measurements[4]. Increasing compression diminishes the amplitude of the transition, that seems to disappear beyond $\approx 11 GPa$, yielding to a metallic state at the highest pressures.

In fact, three anomalies have been previously reported. The first at $87K$, clearly detected by specific heat and magnetic susceptibility measurements[12], but barely seen in the temperature derivative of the electrical resistivity of monocrystals. The second one at $53K$ associated to the development of the insulating state, which has been related to a charge pinning on the subjacent $Na$ cation order[4] or to a magnetic order[13]. The third one, more ill defined in temperature, at $\sim 20K$ has been linked to a second magnetic ordering in the $Co$ spin network. As for all metal - insulator transitions[14], the logarithmic temperature derivative of the electrical resistance determines our transition temperatures. The powder nature of our sample does not allow us to detect the $87K$ anomaly, but we observe the one at $53K$ ($T_c$ in Fig. 2), as a defined peak, as is expected for a second order phase transition. And we can follow with pressure the $20K$ anomaly as a maximum of the logarithmic temperature derivative of the electrical resistance ($T_*$ in Fig. 2). This anomaly does not have the form expected for a phase transition, but that corresponding to a crossover between two different transport regimes, e.g. from conducting to localized. In fact, it is shown on the insert of Fig. 1, where the logarithm of the electrical resistance for each pressure is plotted as a function of $T^{-1/4}$, the good agreement (linear behavior in the graph) of the data and the Mott variable-range-hopping (VRH)[15] law. Thus it seems to be clear that the transition to an insulating state is indeed a crossover from a metallic behavior to a regime where defects localize the carriers remaining below $T_c$.

As shown in Fig. 2, we are able to perform a temperature scaling using the same standard quantum critical pressure dependence of the transition temperature for both anomalies, $|P-P_c|^m$, with $P_c = -28 \pm 5 GPa$ and $m = 0.6 \pm 0.05$. We find that the as-defined transition temperatures for both anomalies increase with pressure (Insert Fig. 2).

Paradoxically, while pressure seems to delocalize carriers, presumably gapped by the successive transitions, and renders the sample metallic, at the same time it increases the transition temperature and its correlated gap. This contrast is visible in the insert Fig. 3, where, together with the transition temperatures, we have plotted the value of the resistance at 10K as a function of pressure. We observe a decrease of the resistance of two orders of magnitude leading to a metallic state, while the transition temperatures are augmented. As in an itinerant picture, the transition temperature is directly related to the number of gapped carriers, an increase of the transition temperature should imply more gapped carriers and a more insulating ground state, not the more metallic ground state that we observe. A prompt and naïve explanation would be to de-correlate the transitions, e.g. of non-itinerant magnetic, and the localization of carriers, though we discuss other interpretations below.

Interestingly, at the higher pressures we obtain a resistivity with a Fermi liquid $T^2$ low temperature dependence (inset Fig. 1), that can imply carrier-carrier or carrier-spin fluctuations scattering. However, some residual localization continues to exist at the lowest temperatures as shown by the slight upturn tail. As similar $T^2$ resistivities have been observed in single crystal samples of the $x = 0.31$ compound, pressure apparently shifts the position of the sample towards the $Na^+$ depleted region of the phase diagram.

b) $Na_{0.31}CoO_2$   We have measured five samples of the same batch of $x = 0.31$ powder. We observe, immediately after pressure application, a time evolution from a more conducting to a less conducting behavior, which stabilizes after several temperature cycles. We show on Fig. 3 this evolution for two of the samples. After pressure stabilization we do not observe the metallic temperature dependence previously reported for single crystal measurements along planes. Damaged grain surfaces, or other powder induced anomalies such as large intergrain resistances, critical in a sample made out of compacted grains with random orientations, may produce this effect. Notwithstanding, it is possible that this behavior may be totally a consequence of our high initial pressure. Compression increases the interaction between the cobalt oxide layers and the $Na^+$ ions. The evolution can thus be attributed to the strong interaction to the $Na^+$ ions with the cobalt lattice, that finally blocks down the ions in a position that pins the carriers, inducing an insulating state on the ambient pressure metallic state. The non-observation of a metallic regime can be the result of the impossibility of measuring the fine powder sample at ambient or sufficiently low pressure. On the insert of Fig. 4, we show the variation of the electrical resistance temperature dependence with pressure for the $x = 0.31$ concentration. The search for transition induced anomalies through the logarithmic derivative, yields once more a transition under pressure (we have only measured above $0.75 GPa$ in this sample), similar to the one found on the 0.5 sample, as shown in Fig. 4 for two different samples. This is certainly a pressure induced effect[1], as no such anomaly has been observed in single crystal measurements at ambient pressure, only a light change in the anisotropy [16]. Though the value and pressure variation are similar to those of the $x = 0.5$ sample (see insert Fig. 3), the powder nature of our sample, that ensures ideal sodium homogeneity, discards any possible contamination with $x = 0.5$ grains. Thus, our measurements imply that compression triggers a similar transition in $x = 0.31$ powder samples.

c) $Na_{0.75}CoO_2$   On Fig. 5 we show the resistance as a function of temperature for different pressures on a $x = 0.75$ sample. At low pressures we observe a trend towards a more metallic character, followed by a gradual upturn at low temperatures that at the highest pressure starts at ambient temperature. As in previous reports on powder samples[17], we do not observe the transition at 20K that signals magnetic ordering. Low pressure ($P \leq 1 GPa$) susceptibility measurements[18] on single crystals of this sodium concentration have shown a very strong increase of this transition with pressure ($dT_c/dP \approx 4 K \cdot GPa^{-1}$). Contrary to the $Na_{0.5}CoO_2$ case, we start here with a metallic state and end up with an almost insulating state at the highest pressure, implying a strong carrier localization with compression, for the highest pressures starting even at room temperature. We remark that a similar, but much less marked, behavior has been observed in the resistivity as a function of $Na$ concentration. In the inset of Fig. 5 we show the comparison of the two cases at two different temperatures, from where we estimate that the effect of $10 GPa$ would be equivalent to about 0.1 in $Na$ concentration.

We begin the discussion by the apparent equivalency between pressure and $Na$ concentration. Charge transfer under compression is a common concept in pressure measurements, as found in, e.g. cuprates[19] or manganites[20]. An instability towards slightly different valencies of non-equivalent cobalt atoms, has been suggested theoretically[9,21] and observed experimentally[4,5]. As discussed in Ref. 17, $Na^+$ ordering should contribute to this as a macroscopic occupation of the $Na(1)$ site[22], being directly above or below $Co$ ions, presumably decreases their valence with respect to the other $Co$ ions, that just surround $Na(2)$ sites. Pressure obviously increases this effect through a stronger compression of the $c$ axis with respect to the $a$ axis[23]. The extra charge localized by compression on the $Co$ above (below) the $Na(1)$ sites, depletes the $a_g$ band, as does a reduction of $Na$ concentration. Thus it seems reasonable to observe that application of pressure is equivalent to a $Na$ concentration decrease. This parallellism between pressure and $Na$ concentration variation can be also expected from band structure calculations that have shown that the oxygen $z$ coordinate, half the thickness of the $CoO_2$ layer, is quadratically proportional to the $Na$ concentration. As pressure compresses the $z$ axis, we can expect as a consequence a change in the density of states at the Fermi level, that can be express as a change in $Na$ concentration . We can estimate an order of magnitude value for this change, $dx/dP$. From Zhang et al. , we obtain $dx/dz \approx 7 Å^{-1}$ at $x = 0.75$, and using the measured $c$ parameter compressibility[24], $\beta_c = 0.004 GPa^{-1}$, we find $dx/dP \approx 0.026 GPa^{-1}$, in agreement with the variation estimated from the inset of Fig. 5, $dx/dP \approx 0.01 \pm 0.008 GPa^{-1}$. This equivalence can explain both the increase of resistance at low temperatures in the $x = 0.75$ compound and the metallization of the $x = 0.5$ sample. Furthermore, we observe a transition in the $x = 0.31$ sample, as could be predicted from the existence of a transition below $15 GPa$ in the $x = 0.5$ , that should be equivalent to an $x = 0.35 \pm 0.075$ concentration, though the value is lower than the expected one. In fact, according to band structure calculations, nestings equivalents to the one held responsible[25] for the transition in the $x = 0.5$ cobaltite should also be expected in the $x = 0.31$ material. On the other hand, evidence for a reconstruction of the Fermi surface at low temperatures has been reported for $x = 0.3$ single crystals through Shubnikov-de Haas oscillations[26]. We should stress, though, that this equivalence should be limited, as,e.g., steric effects due to the introduction of extra $Na$ ions cannot be mimic by pressure.

The insulating state observed at low temperatures in $Na_{0.5}CoO_2$ has been attributed either to a charge density wave (CDW) or to magnetic order or spin density wave (SDW), but it is, as we shall point out, more complex. The large majority of the CDW/SDW (DW) compounds (1-D, 2-D or even 3-D, for a detailed list see Ref. 27) that have been studied up to date yield a ratio gap $\Delta$ to transition temperature $f = \Delta/(k_B T_{c1})$ much larger that the theoretical mean-field value, i.e. $f \approx 1.75$, going up to almost 10. Such large ratios have been attributed to phonon entropy[28], to inelastic scattering of carriers by phonons[22] or to deviations from perfect nesting in the SDW case[29,30]. The different and degenerate possibilities of nesting allow the formation of short range order (SRO) DW fluctuations at a temperature $T_{MF}$ with the expected mean field gap, but that have no long range order (LRO)[31]. The transition temperature observed in transport properties just signals the development of LRO at a temperature $T_{LRO}$ much smaller than $T_{MF}$. The optical measurements[32,33] done on cobaltite samples compared to the transition temperature measured in transport properties yield $f \approx 1.75$, the mean field value, which should be stressed is highly unusual. To obtain a mean field value of $f$, external effects, such as commensurate pinning, are obligatory to stabilize the SRO fluctuations and allow the development of LRO near to $T_{MF}$. We can thus speculate that the underlying $Na^+$ order may play an essential role in the appearance of the transition at $T_{c1}$, that corresponds to $T_{LRO}$, by stabilizing the DW (even under the assumption of an SDW, a small lattice distortion appears at $T_{c1}$[34]) through matching pinning or Fermi surface reconstruction[25]. Compression will increase the interaction between the $Na^+$ order and the DW, probably leading to the increase of the transition temperatures that we observe with pressure.

In conclusion, we show that pressure enhances the low temperature transition temperatures in the $x=0.5$ compound, while the insulating ground state disappears under pressure. This result contradicts the present understanding of DW's theory and its application to layered cobalt oxides. We speculate that the increase of the transition temperature is due to an enhancement of the ordered state due to stronger pinning by the $Na^+$ sublattice, and that the passage to a metallic ground state is due to depletion of the $a_g$ band under pressure by electron pinning on $Na(1)$ sites. Furthermore, the observation of a transition for the $x=0.31$ samples under pressure should encourage its search with other methods in single crystal samples.

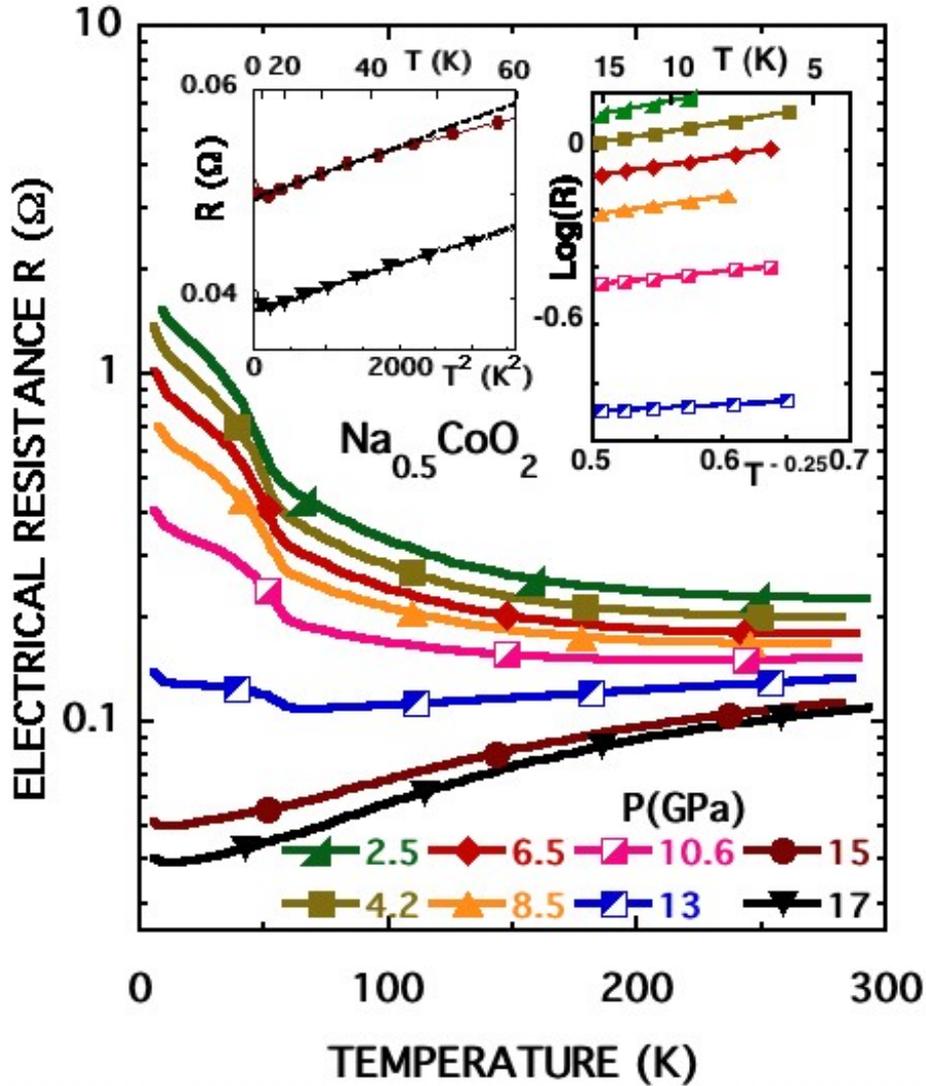

**Figure 1 :** (Color online) Temperature dependence of the electrical resistance of the $Na_{0.5}CoO_2$ sample D at different pressures. The transition towards an insulating ground state is clearly seen below $11 GPa$. Above this pressure the behavior is metallic. Left inset: Detail showing the $T^2$ at low temperatures for the highest pressures, where the samples are metallic state (albeit a small residual localization at very low temperatures). Right inset: plot showing the 3D VRH behavior of the low temperature electrical resistance.

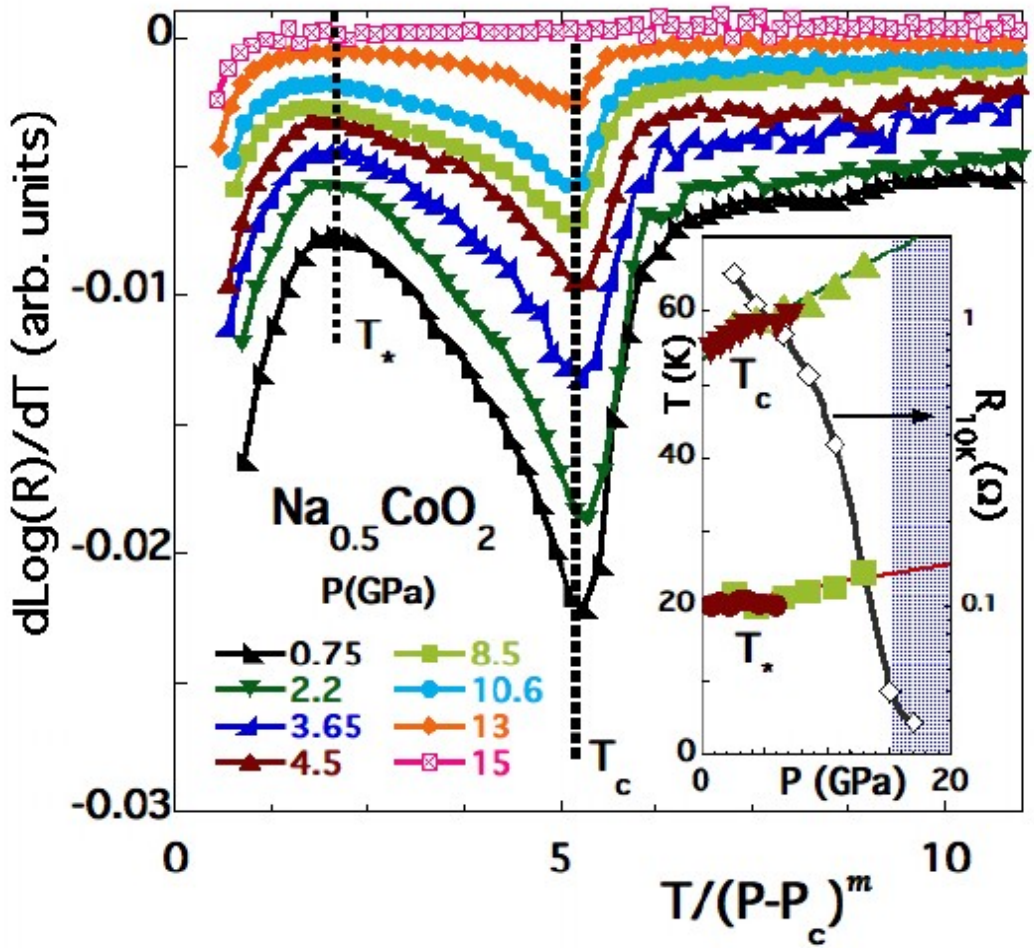

**Figure 2 : (Color online) Logarithmic temperature derivative of the resistance of two $Na_{0.5}CoO_2$ samples (triangular symbols : sample C; other symbols sample D) as function of temperature scaled by $|P-P_c|^m$, with $P_c = -28 \pm 5 GPa$ and $m = 0.6 \pm 0.05$. Insert: Pressure dependence of the low temperature transition temperatures $T_c$ (triangles) and $T_*$ (dots and squares) for two $Na_{0.5}CoO_2$ samples (dark symbols : sample C; light symbols sample D). The solid lines represent the scaling dependences $T_{ci}^0 \cdot |P-P_c|^m$. We also show the two orders of magnitude fall of the $10K$ resistance ($R_{10K}$, white diamonds), that occurs while the transition temperatures increase with pressure. The region in pressure where the samples have a metallic behavior is shaded.**

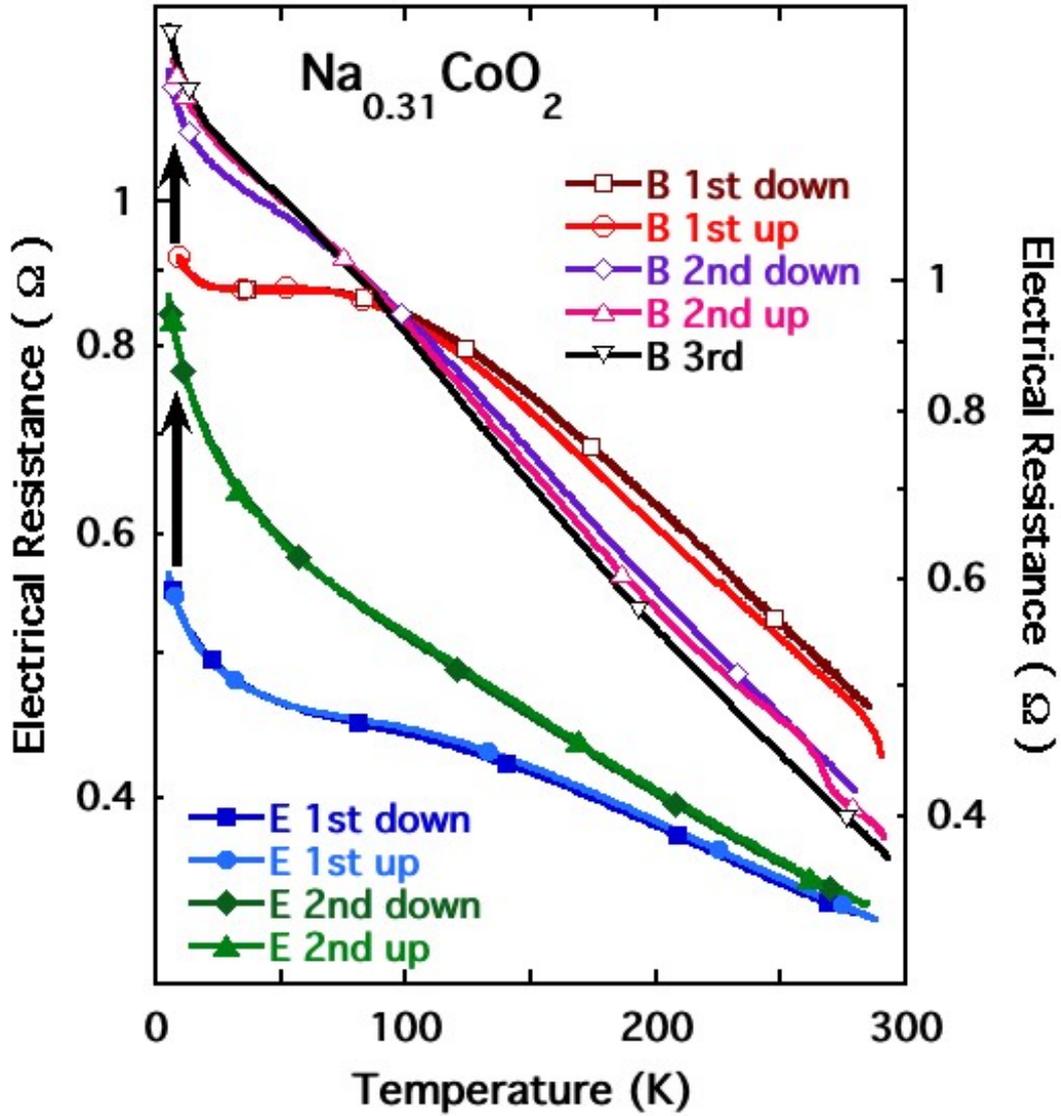

**Figure 3 (Color Online) Time evolution at fixed pressure of the teperature dependence of two $Na_{0.31}CoO_2$ samples, B (P=0.75GPa) and E (P=2GPa). The effect of pressure, together to the cycling in temperature, increases the resistance of the sample.**

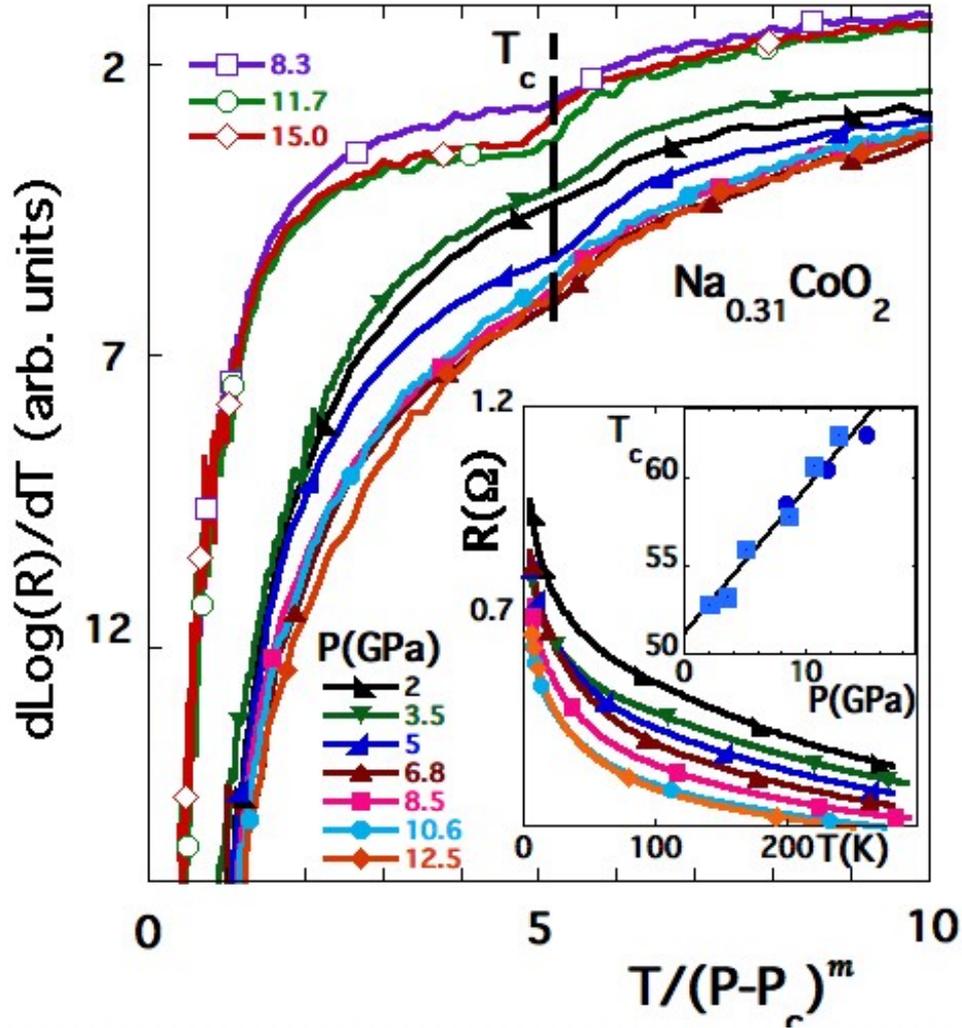

**Figure 4 : (Color online) Logarithmic temperature derivative of the resistance of two $Na_{0.31}CoO_2$ samples (empty symbols : sample A; full symbols sample F) as function of temperature scaled by $|P-P_c|^m$, with $P_c = -31 \pm 3 GPa$ and $m = 0.6 \pm 0.05$. Insert: Temperature dependence of the electrical resistance of the $Na_{0.31}CoO_2$ sample F at different pressures. Inner insert: Pressure dependence of the transition temperatures $T_c$ for two $Na_{0.31}CoO_2$ samples (circles : sample A; squares sample F). The solid line represents the scaling dependences $T_{ci}^0 \cdot |P-P_c|^m$.**

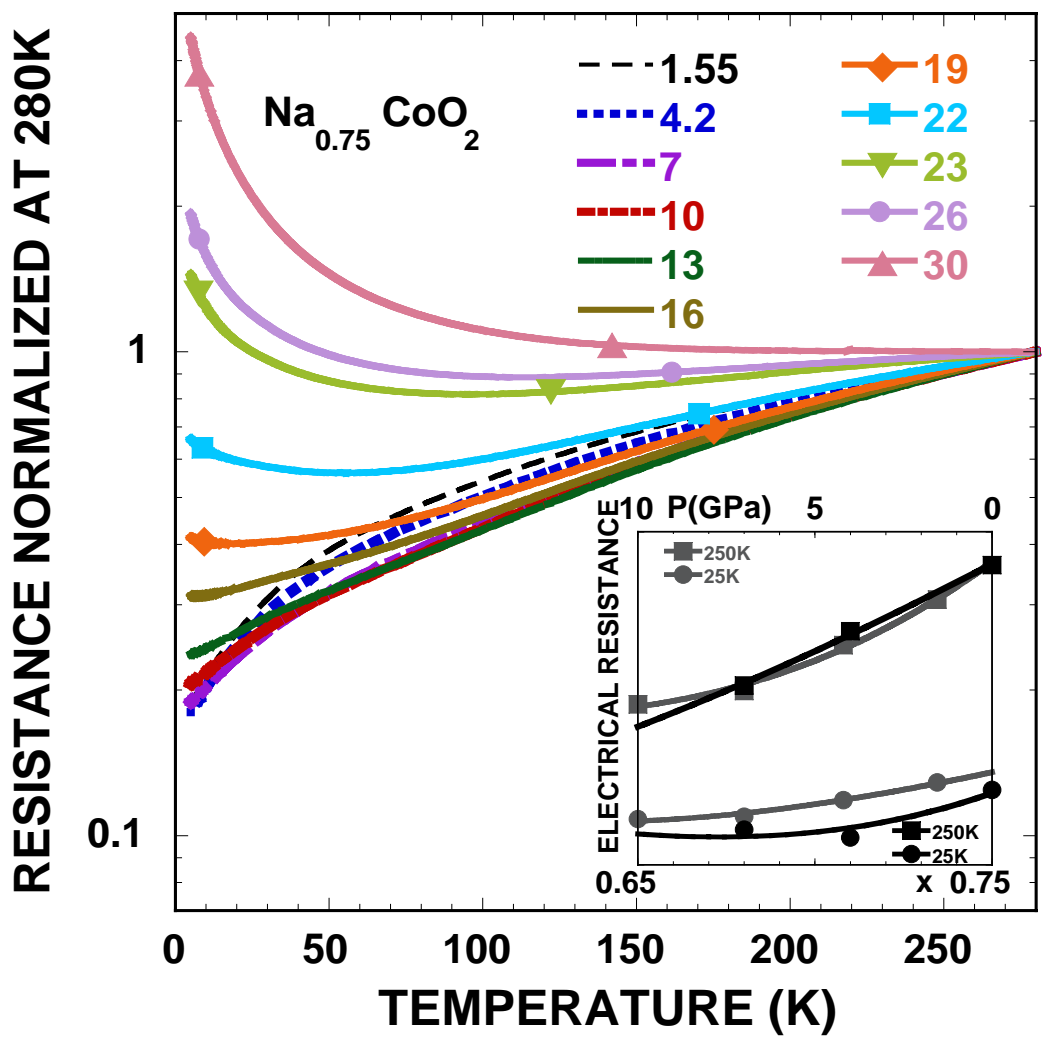

Figure 5 : (Color online) Temperature dependence of the electrical resistance of two $Na_{0.75}CoO_2$ samples normalized at 280K (for clarity convenience) for different pressures. Inset : Comparison of the values of normalized resistance for 250K and 25K as a function of pressure (gray symbols; this work) and of sodium concentration (black symbols; from Ref. 4).